\documentclass[a4paper,11pt]{article}
\usepackage{pos}
\usepackage{graphicx}
\usepackage{bm}
\usepackage{subcaption}
\usepackage{hyperref}

\title{From scattering towards multi-hadron weak decays}
\author[]{Felix Erben}

\affiliation[]{CERN, Theoretical Physics Department, Geneva, Switzerland}

\emailAdd{felix.erben@cern.ch}

\abstract{In this article I provide an overview of the current state of scattering within lattice QCD, along with ongoing projects that examine weak decays involving scattering states as either final or intermediate states. Significant progress has been made in the study of multi-hadron weak decays, opening the door for scattering calculations to make meaningful contributions to flavour physics and further establishing lattice QCD as the key non-perturbative tool for QCD predictions. In addition to discussing new calculations, I also highlight recent advancements in finite-volume formalisms, which enable the exploration of previously inaccessible channels.}

\FullConference{The 41st International Symposium on Lattice Field Theory (LATTICE2024)\\
 28 July - 3 August 2024\\
Liverpool, UK\\}

\begin{document}
\maketitle

\section{Resonances on the lattice}

Lattice QCD computations employ a Euclidean path integral formulation, using Euclidean time $t$. Consequently, correlation functions calculated in lattice QCD do not exhibit real-time behaviour. However, at sufficiently large Euclidean time $t$, matrix elements $\langle 0 | \hat{O} | n \rangle$ and finite-volume energy states $E_n$ can be determined by fitting to the spectral decomposition:
\begin{align}
\langle O_1(t) O_2(0) \rangle = \sum_n \frac{1}{2E_n}\langle 0 | \hat{O}_1 | n \rangle \langle n | \hat{O}_2 | 0 \rangle e^{-t E_n} \, .
\end{align} 
Importantly, while these matrix elements and energies are derived from Euclidean correlation functions, they contain no dependence on the metric signature and could equally well have been determined from Minkowski signature correlation functions.

\begin{figure}
    \centering
    \includegraphics[width=6cm]{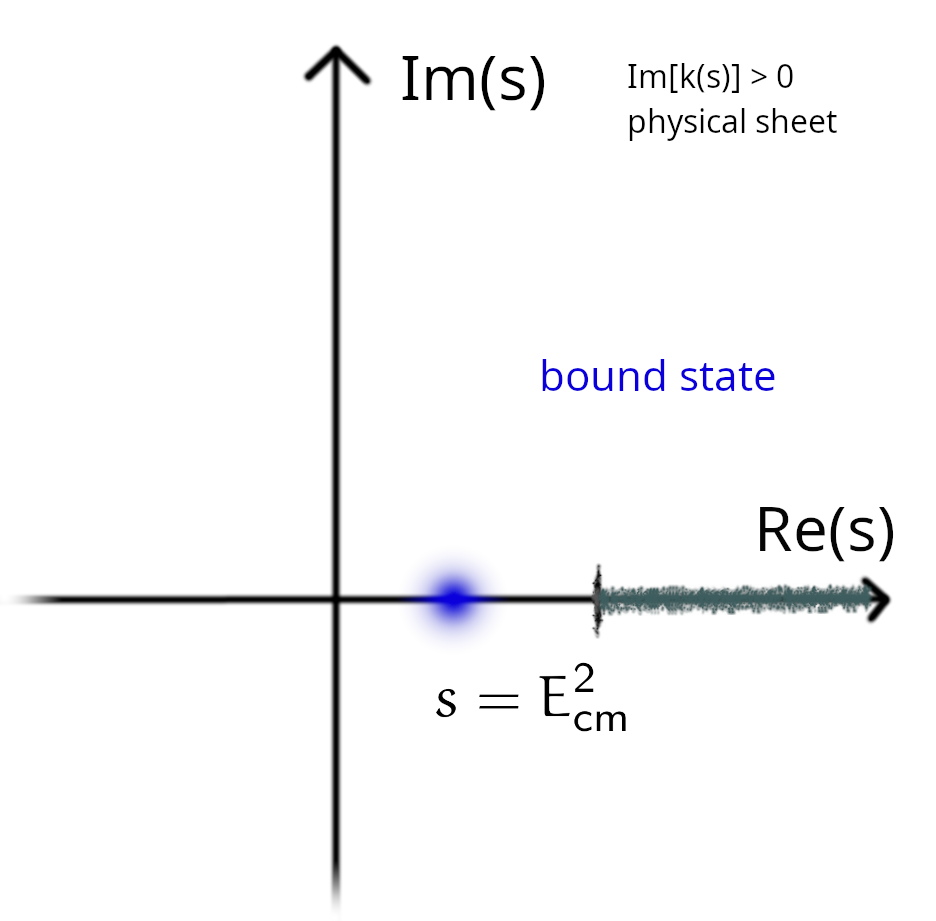}
    \includegraphics[width=6cm]{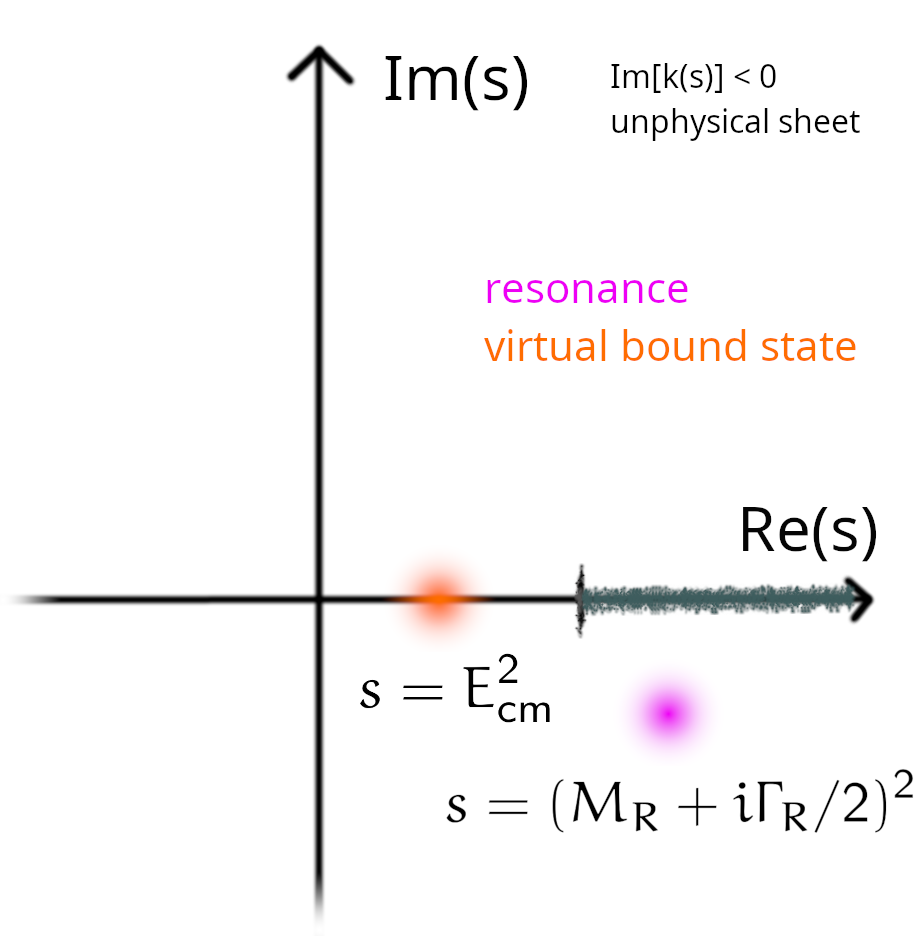}
    \caption{Complex plane of the Mandelstam variable $s$, illustrating possible pole singularities of the scattering amplitude. Sheets are distinguished by the imaginary part of $k$. The left plot shows the physical sheet where bound states occur on the real axis. The right plot shows the unphysical sheet where resonances appear above the branch cut and bound states may manifest as virtual bound states.}
    \label{fig:s-plane}
\end{figure}

The approach described above effectively captures behaviour asymptotic in Euclidean time $t$, rendering it applicable for studying QCD-stable hadrons such as pions, kaons, and protons. Conversely, resonances are unstable QCD states decaying into multiple stable hadrons. In infinite-volume, continuum QCD, a resonance can be described as a pole singularity in the scattering amplitude, which can be determined in the complex Mandelstam $s$ plane using the scattering amplitude
\begin{align}
t^\ell(\sqrt{s}) = \frac{1}{\cot \delta^\ell(\sqrt{s}) - i} \, ,
\label{eq:pole-continuation}
\end{align}
where $\ell$ denotes the definite orbital angular momentum or partial wave of the infinite-volume scattering phase shift $\delta^\ell(\sqrt{s})$. Examining the complex $s$ plane reveals that for a single channel $\ell$, the square-root cut generates two Riemann sheets. Each additional channel introduces another cut, effectively doubling the number of Riemann sheets. The resonance pole position $s = (M_R + i \Gamma_R/2)^2$ with mass $M_R$ and width $\Gamma_R$ lies above the cut and away from the real axis. As sketched in Fig.~\ref{fig:s-plane}, bound states can also occur on the real axis. These typically appear on the physical sheet where $\mathrm{Im}[k(s)] > 0$. Here $k$ is defined by the relation $\sqrt{s}=\sqrt{k^2 + m_1^2} + \sqrt{k^2 + m_2^2}$ for scattering of two particles with masses $m_1$ and $m_2$. Bound states may also appear on the unphysical sheet where $\mathrm{Im}[k(s)] < 0$ as virtual bound states. 

In lattice QCD, the finite volume obscures resonance poles, branch cuts, and multiple Riemann sheets. Instead, a discretized energy spectrum can be calculated. This process involves defining an interpolator basis incorporating all single- and multi-hadron states in the resonance system. For instance, for a mesonic vector resonance like the $K^*(892)$, vector bilinears and two-hadron $K\pi$ interpolators with various meson momenta must be included to have overlap with all low-lying states in the system. After computing correlation functions for all interpolators, solving a generalized eigenvalue problem (GEVP)~\citep{Luscher1990a,Blossier:2009kd,Fischer2020a} yields optimized correlators from which the lowest-lying states of the discretized spectrum can be extracted.

This finite-volume spectrum differs from the spectrum of energies for non-interacting particles. L\"uscher~\citep{Luscher:1985dn,Luscher:1986pf,Luscher:1990ux} demonstrated that these finite-volume energy shifts represent predictable imprints of the finite volume on the spectrum, directly relating it to the infinite-volume scattering phase shift $\delta^\ell(\sqrt{s})$. Originally derived for two identical particles at rest, this formalism has been extended for broader cases~\citep{Rummukainen:1995vs,Kim:2005gf,Fu:2011xz,Leskovec:2012gb,Bernard:2010fp,Doring:2011vk,Briceno:2012yi,Hansen:2012tf,Briceno:2014oea}. While lattice phase shift predictions occur at discrete energies $\sqrt{s} = E_\mathrm{cm}$, a continuous description of the phase shift can be achieved through fitting of a model to this data. See Refs.~\citep{Briceno2018,Mai:2022eur} for recent reviews.

The structure of the remainder of this document is as follows: In Section~\ref{sec:recent}, I will review recent developments in scattering computations and advancements in formalism. Section~\ref{sec:our-phys-comp} focuses on a calculation of the $\rho$ and $K^*$ resonances at physical pion masses, a project I co-authored. The review then progresses with a discussion of works related to processes involving electroweak matrix elements in Section~\ref{sec:eweak}. Finally, I will conclude with a brief outlook in Section~\ref{sec:outlook}. 

\section{Recent scattering works}
\label{sec:recent}

A clear indication of the growing importance of scattering in lattice QCD is that Lattice 24 featured two separate plenary talks on the subject. This allowed me to delve deeply into many of the recent exciting developments, while the other presentation focused on progress related to exotic states. For a more comprehensive overview of recent advancements in scattering, I encourage readers to also read N. Mathur's proceedings~\citep{Lattice24:mathur}. For works that precede the ones covered in these proceedings the reader is refered to A. Hanlon's proceedings~\citep{Hanlon:2024fjd} on scattering for the Lattice 23 conference. I will also not cover works that have first appeared after my plenary talk. 

\subsection{Various two-particle scattering works}

At the $SU(3)_F$ symmetric point, where the light and strange quark masses are equal, i.e., $m_u=m_d=m_s$, a study was conducted on elastic S-wave scattering of a charmed meson with a light pseudoscalar meson ($\pi/K$) in the $J^P=0^+$ channel~\citep{Yeo:2024chk}. The analysis employed three lattice volumes with $M_\pi \sim 700$ MeV. The high symmetry present at the $SU(3)_F$ point results in three relevant sectors in the flavour representation: $\mathbf{\overline{3}}, \mathbf{6}, \mathbf{\overline{15}}$. Utilising an extensive basis of interpolating operators, the authors identified a deeply bound state corresponding to the $D^*_{s0}(2317)$ in the $\mathbf{\overline{3}}$ sector. Additionally, a virtual bound state was observed in the $\mathbf{6}$ sector, located within the region of $2510 - 2610$ MeV. While direct conclusions regarding physical-point scattering cannot be easily drawn from this analysis, such studies offer valuable insights into the influence of flavour symmetries on particle dynamics.

Another study investigated $D\pi$ scattering away from this symmetry point~\citep{Yan:2024yuq}. The authors explored several pion masses ranging from $132$ MeV to $317$ MeV which means that they, for the first time in this channel, included physical pion masses in their analysis. However, at the physical point, the results exhibit very large uncertainties. By comparing their findings with previous works in this channel~\citep{Mohler:2012na,Moir:2016srx,Gayer:2021xzv}, the pion mass dependence of the pole position was mapped out. The results confirm that (virtual) bound states appear in the channel for $M_\pi \gtrsim 300$ MeV, while resonances are observed for $M_\pi \lesssim 270$ MeV.

\begin{figure}
    \centering
    \includegraphics[width=6cm]{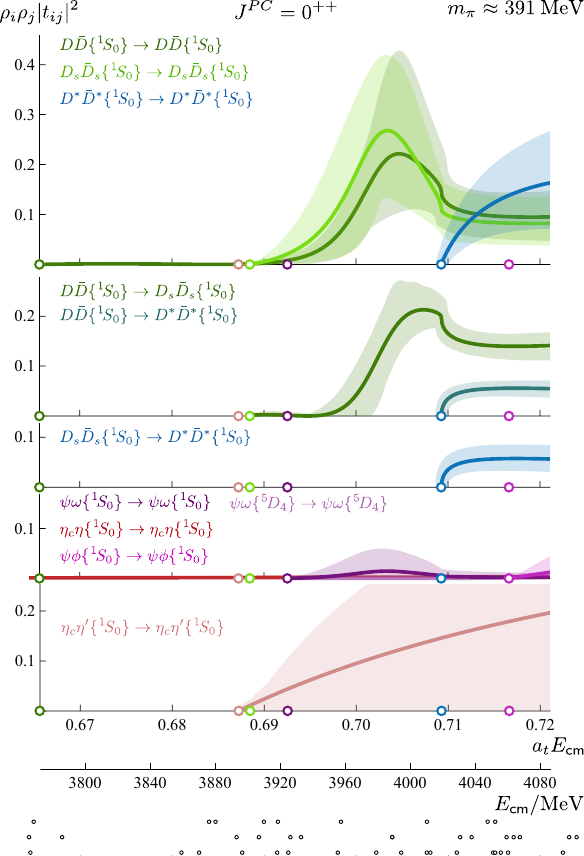}
    \ \ \ \ \ \ \ \ \ \ 
    \includegraphics[width=6cm]{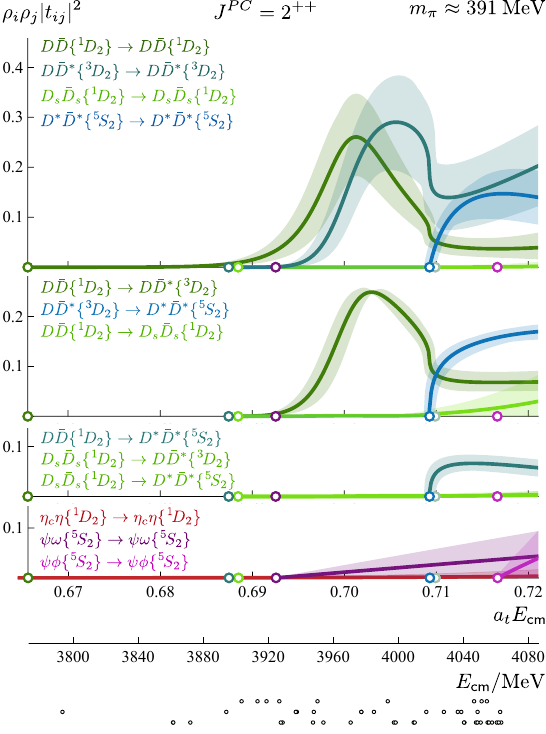}
    \caption{Scattering amplitudes for the charmonium resonances computed in~\citep{Wilson:2023anv,Wilson:2023hzu}. The left plot shows the $J^{PC}=0^{++}$ and the right plot shows the $J^{PC}=2^{++}$ channel. Shaded bands are the envelope over fit variations to the energy levels shown as black dots at the bottom of the plots.}
    \label{fig:hadspec-charmonium}
\end{figure}

Furthermore, a computation has appeared of $\chi_{c0},\chi_{c2}$ charmonium resonances~\citep{Wilson:2023anv,Wilson:2023hzu}. These channels are interesting due to experimental discoveries and puzzles like the $X(3872)$ at Belle~\citep{Belle:2003nnu} or the $Z_c(3900)$ at BESIII~\citep{BESIII:2013ris} and Belle~\citep{Belle:2013yex}. Despite challenges due to the very high density of scattering channels, the lattice study found that the L\"uscher method with its extensions continued to work well, and found interactions in quark-disconnected channels (like $J/\Psi - \omega$) much suppressed when compared to interactions in quark-connected channels (like $D\bar{D},D_s\bar{D}_s$). They found a single resonance pole in $J^{PC}$ channels $0^{++}$ and $2^{++}$ coupled to multiple channels, shown in Fig.~\ref{fig:hadspec-charmonium}. Two extra states so far unobserved (but not unexpected) were found in $2^{-+}$ and $3^{++}$.

\subsection{The tetraquark $T_{cc}^+(3875)$}

The tetraquark $T_{cc}^+(3875)$, an exotic state, was discovered by LHCb in 2022~\citep{LHCb:2021vvq}. It is positioned very close to the $DD^*$ threshold:
\begin{align}
    M_{T^+_{cc}} - (M_{D^{*+}}+M_{D^0}) &= -0.27(6)\, \text{MeV} \, .
\end{align}
The $D^*$ resonance is remarkably narrow, with a width of $\Gamma_{D^{*+}} \sim 80$~keV and a mass of $M_{D^{*+}} \sim 2$~GeV. The sole observed decay mode for the tetraquark is $T_{cc}^+(3875) \to D^0D^0\pi^+$, making it a prime candidate for applying three-particle scattering formalisms~\citep{Hansen:2024ffk}.

For lattice QCD practitioners, this distinct hierarchy of scales suggests that for slightly heavier-than-physical pions $M_\pi \gtrsim 150$ MeV, the $D^*$ becomes a stable resonance, effectively approximated by a vector-like bilinear interpolator using the narrow-width approximation. A first computation using this approach appeared in 2022~\citep{Padmanath:2022cvl} and was soon followed by other studies~\citep{Chen:2022vpo,Lyu:2023xro}. Notably, the pole position of the $T_{cc}$ lies near the left-hand branch cut associated with one-pion exchange, potentially leading to significant finite-volume effects not accounted for in the L\"uscher formalism. This challenge was first encountered earlier in the H-dibaryon spectrum~\citep{Green:2021qol}.

\begin{figure}
    \begin{subfigure}{0.48\textwidth}
        \includegraphics[width=\textwidth]{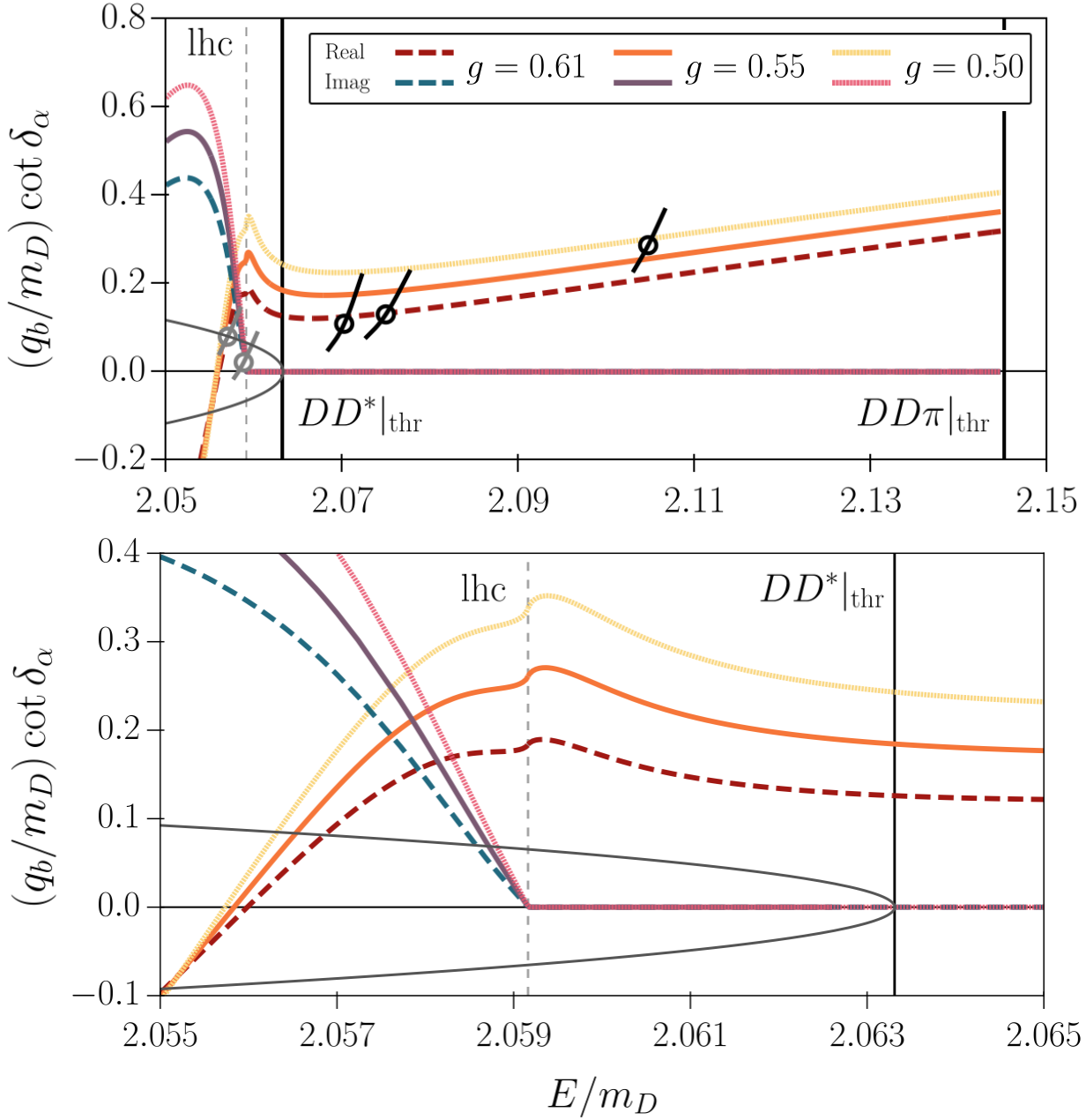} 
    \caption{Three-particle solution from~\citep{Dawid:2024dgy} which they comapre to the data of~\citep{Padmanath:2022cvl}. Shown is $q \cot\delta$ as a function of energy, with the bottom plot a zoom-in of the top one. Crucially, no pole appears left of the $DD^*$ threshold.}
    \label{fig:3p-lhc}
    \end{subfigure}
        \hspace{0.1cm}
     \begin{subfigure}{0.48\textwidth}
    \centering
    \includegraphics[width=\textwidth]{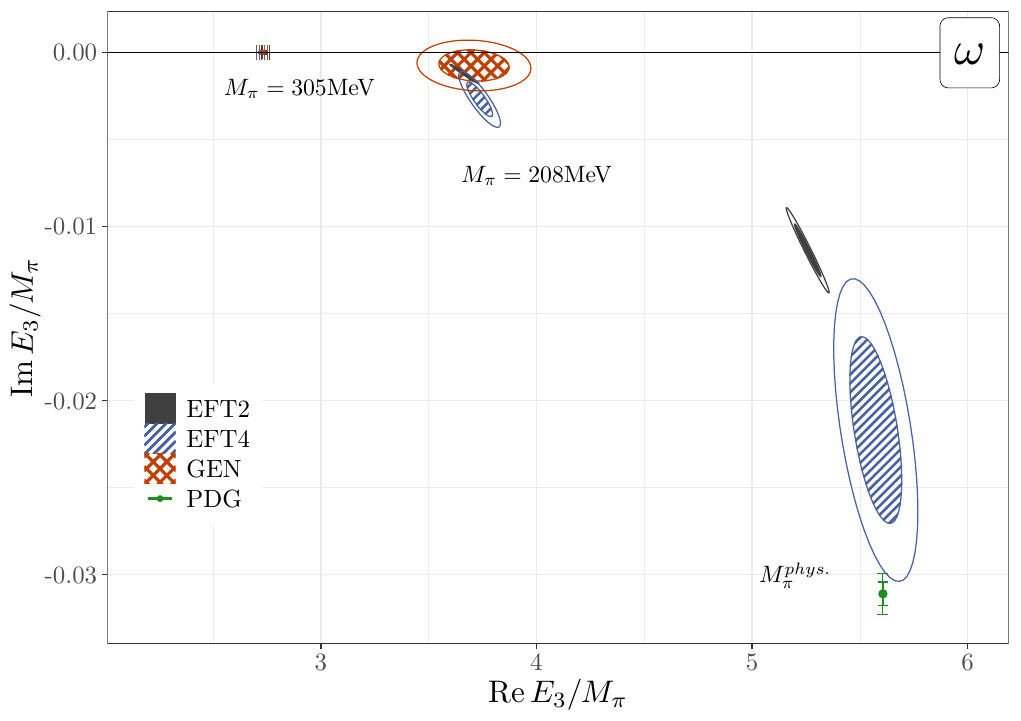}
    \vspace{0.9cm}
    \caption{Pole position of the $\omega$ meson calculated in~\citep{Yan:2024gwp}. Extrapolations of the two calculated pole position using an effective field theory inspired method (EFT4) overlap with the experimental result. }
    \label{fig:omega-pole}
    \end{subfigure}
    \caption{Plots from the $T_{cc}$ and $\omega$ works involving 3-particle formalisms. }
\end{figure}

This issue has prompted refinements of the L\"uscher method to account for the left-hand cut explicitly. One solution~\citep{Raposo:2023oru} modifies the formalism by more carefully projecting onto on-shell intermediate states, yielding a modified quantization condition involving a K-matrix free of the cut. The physical scattering amplitude is then reconstructed using integral equations. An alternative approach~\citep{Bubna:2024izx}, also presented at the conference, similarly separates long-range and short-range components in the derivation and arrives at an analogue of the intermediate K-matrix. In this second approach, the K-matrix can be related to the physical scattering amplitude via an algebraic relation, so the necessity to solve integral equations is avoided. Yet another approach tailored to the $T_{cc}$~\citep{Hansen:2024ffk} uses a three-particle formalism that sidesteps the left-hand cut issue by allowing pion exchange between the $D$ and $D^*$. This approach has been numerically implemented~\citep{Dawid:2024dgy}, showing good agreement with lattice data~\citep{Padmanath:2022cvl} and phase shifts from the three-particle formalism. Their plot comparing their solution to lattice data is shown in Fig.~\ref{fig:3p-lhc}.

Additional strategies to address the issue include incorporating a local diquark-antidiquark operator~\citep{Vujmilovic:2024snz}, which modestly affects the pole position, enhancing the system's attraction. This effect is more pronounced for the $b$-quark system with the $T_{bb}$ tetraquark. Another study~\citep{Collins:2024sfi}, conducted at $M_\pi = 280$ MeV across five charm-quark masses bracketing $m_c^\mathrm{phys}$, employed an S-wave fit to the Lippman-Schwinger equation with an EFT potential that incorporates the left-hand cut. At the physical charm-quark mass they find a resonance pole, that becomes a bound state for higher heavy-quark mass. A similar transition from resonance to bound state was known from earlier studies on ensembles with smaller light-quark masses but at the physical charm-quark mass~\citep{Padmanath:2022cvl,Chen:2022vpo,Lyu:2023xro}.

Further insights into the $T_{cc}$ were gained from a coupled-channel $DD^*, D^*D^*$ scattering analysis~\citep{Whyte:2024ihh}. This study identified two poles in the S-wave: a virtual bound state consistent with the $T_{cc}$ found in single-channel analyzes and a resonance below the $D^*D^*$ threshold, referred to as $T'_{cc}$, which primarily couples to the $D^*D^*$ S-wave channel but also influences the $DD^*$ S-wave amplitude.

\subsection{Three-particle scattering}

Finite-volume scattering formalisms were initially developed for systems involving two particles. However, several formalisms for three-particle scattering have been available for some time now \citep{Hansen:2014eka,Hansen:2015zga,Hammer:2017uqm,Hammer:2017kms,Mai:2017bge,Briceno:2018aml}. For a detailed discussion and comparison of the three available approaches, which are mutually compatible where a comparison is applicable, the reader is referred to recent reviews on three-particle formalisms \citep{Hansen:2019nir,Mai:2021lwb}.

Recent formal advancements presented at the conference included an extension of the three-particle formalism to multiple channels of distinct particles, illustrated using the $\eta\pi\pi + K\overline{K}\pi$ system in isosymmetric QCD \citep{Draper:2024qeh}. Additionally, the first implementation of the previously derived~\citep{Draper:2023xvu} three-neutron quantization condition was presented \citep{Schaaf:2024qer}, marking a significant step towards a deeper understanding of three-nucleon forces and ultimately contributing insights into the properties of neutron-rich nuclei. Furthermore, a first implementation of the previously derived relativistic-field-theory finite-volume formalism across all three-pion isospin channels~\citep{Alotaibi:2024fyt} was introduced. Previously, this quantization condition had only been implemented for maximal isospin, corresponding to the case of three identical $\pi^+$ mesons.

Numerical implementations of three-particle scattering studies were also presented for various three-meson scattering amplitudes, including one work calculating $KKK$, $KK\pi$, $K\pi\pi$, and $\pi\pi\pi$, all at physical quark masses and maximal isospin (where all particles carry positive charge)~\citep{Lattice24:Lopez}. Additionally, the first calculation of the $\omega(782)$ meson, a three-particle resonance in the isoscalar channel, was reported~\citep{Yan:2024gwp}. The $\omega$ meson primarily decays to three pions ($\omega \to 3\pi$), though two of these pions can couple to the $\rho$ meson, making $\omega \to \rho\pi$ another decay channel that needs to be accounted for explicitly. This work was conducted on four CLQCD ensembles, using a single lattice spacing and two different volumes at pion masses of $M_\pi = 210$ MeV and $M_\pi = 300$ MeV. The pole positions of the $\omega$ were determined on all ensembles using various EFT-inspired fits, with extrapolations to the physical pion mass compared to experimental values - shown in Fig.~\ref{fig:omega-pole}. Good agreement was found for one of the fit forms. The study also included an analysis of the $\rho \to 2\pi$ channel, and the $\omega - \rho$ mass splitting was determined to be $29(15)$ MeV.

\subsection{Physical-point $\rho$ and $K^*$ resonances}
\label{sec:our-phys-comp}

A recently published study (of which I am a co-author) focused on the $\rho(770)$ and $K^*(892)$ mesonic resonances \citep{Boyle:2024grr,Boyle:2024hvv}. While these resonances have been extensively investigated in the past, our work's calculations were performed using light and strange quarks at physical masses, albeit limited to a single lattice spacing. Additionally, we developed an extension to data-driven formalisms, enabling us to report both statistical and systematic uncertainties for the final resonance pole positions. All correlator data generated for this study has been made publicly available \citep{boyle_2024_vy9x7-bzn92} as a 760 GB dataset.

The data production employed the distillation method \citep{Peardon:2009gh,Morningstar:2011ka} and has been performed using the Grid \citep{Boyle2015} and Hadrons \citep{Hadrons2023} software frameworks. These tools, developed in part by the authors of \citep{Boyle:2024grr,Boyle:2024hvv}, are open-source and thoroughly documented \citep{Hadrons-doc}. Distillation allows one to effectively compute all two-point correlation functions from a basis of interpolating operators, including vector bilinears:
\begin{equation}
O_V(x) \sim \bar{q}_V(x)\gamma q_V'(x), \quad V \in {\rho^+, K^{*+}},
\end{equation}
as well as two-bilinear operators:
\begin{equation}
O_{MM'}(x,y) \sim \bar{q}_1(x)\gamma_5 q_2(x) \ \bar{q}_1'(y)\gamma_5 q_2'(y), \quad MM' \in {\pi \pi, K \pi}\, ,
\end{equation}
where each meson is projected to individual definite spatial momenta. Because the lattice has reduced cubic symmetry, correlation functions are projected into irreducible representations (irreps) of the cubic group, denoted by $\Lambda$. Correlators in all irreps are assembled into matrices:
\begin{align}
C^\Lambda_{AB}(t,t') = \langle O^\Lambda_A(t) O^\Lambda_B(t')^\dagger \rangle \, ,
\end{align}
where $A,B$ are generic labels for $V, MM'$. In our setup we compute the correlation functions from all $N_T=96$ source times $t'$ and define
\begin{align}
C^\Lambda_{AB}(t) = \frac{1}{N_T}\sum_{t'} C^\Lambda_{AB}(t-t',t')  \, .
\end{align}
From these matrices, effective masses and energy levels of the finite-volume spectrum in the irrep $\Lambda$ can be extracted by solving the generalized eigenvalue problem (GEVP)~\citep{Luscher1990a,Blossier:2009kd,Fischer2020a}
\begin{align}
    C^\Lambda(t) u^\Lambda_n(t)  = \lambda^\Lambda_n(t) C^\Lambda(t_0) u^\Lambda_n(t)  \, ,
\end{align}
This results in $n_{\sf op}$ generalized eigenvalues $\lambda_n(t)$ and eigenvectors $u_n(t)$, with mild dependence on $t_0$. The eigenvalues $\lambda^\Lambda_n(t)$ correspond to optimized interpolators, which couple effectively to state $n$ and allow extraction of both ground and excited states.

To simplify notation, we omit the irrep index $\Lambda$ in the following discussion. Correlation functions are computed across five momentum frames, with lattice momenta ranging from ${0 \leq \mathbf{P}^2 \leq 4 (2 \pi/L)^2}$. Within these frames, we consider all irreps coupling exclusively to the $\ell=1$ or $P$-wave in the partial wave expansion.

\begin{figure}
    \centering
    \includegraphics[width=9cm]{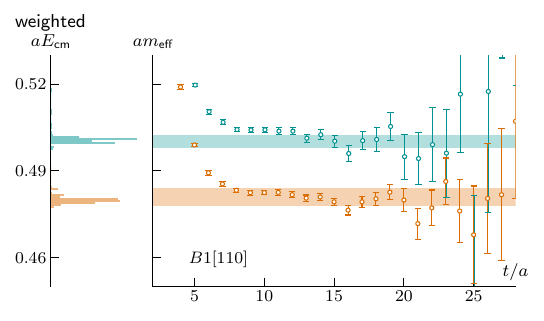}
    \caption{Effective masses of the two lowest GEVP eigenvectors in the ${\bf P}=[110]$ frame, $B_1$ irrep in the $K \pi$ scattering system. The overlaid bands correspond to our fit result for the combined systematic and statistical error, estimated from the weighted histogram plotted at the left-hand side of the plot.}
    \label{fig:histo-effmass}
\end{figure}

The effective masses for the two lowest states in a selected irrep are shown in Fig.~\ref{fig:histo-effmass}. Each level's estimator, which covers statsitical and systematic uncertainty, is determined from a weighted histogram of fit results within reasonable bounds, described further in \cite{Boyle:2024grr}. The weight associated with each fit result entering the histogram is defined by:
\begin{align}
w_{\sf corr} = e^{-\frac 12 {\sf AIC}_{\sf corr}}
\end{align}
where the Akaike information criterion (AIC)~\citep{Akaike1974,Akaike1978} is given by:
\begin{align}
{\sf AIC}_{\sf corr} = \chi^2 - 2n^{\sf par} - n^{\sf data} \, ,
\end{align}
for a fit to $n^{\sf data}$ points with $n^{\sf par}$ parameters and chi-squared $\chi^2$. This data-driven approach has been recently employed \citep{Borsanyi2021} to systematically determine errors on fit-range choices with minimal human intervention.
Due to the structure of the L\"uscher formalism, extracting scattering information requires using the computed energy levels as input for a second fit to a phase shift model. We employed both Breit-Wigner and effective range expansion models for this purpose. Performing a phase shift fit for a single sample of energy levels (i.e. one estimator per energy level) allows straightforward statistical error propagation via the bootstrap method. However, propagating systematic errors is more complex: In an idealistic method, all combinations of underlying correlator fits would be used, weighted by a combined AIC score to obtain the final result. Given approximately $O(20)$ energy levels, each estimated by a histogram over $O(100)$ correlator fits, the number of phase shift fits in this ideal method would be $n_{\sf ideal} \sim 100^{20}$, making direct computation infeasible.

\begin{figure}
    \centering
    \begin{subfigure}{0.58\textwidth}
    \includegraphics[width=\textwidth]{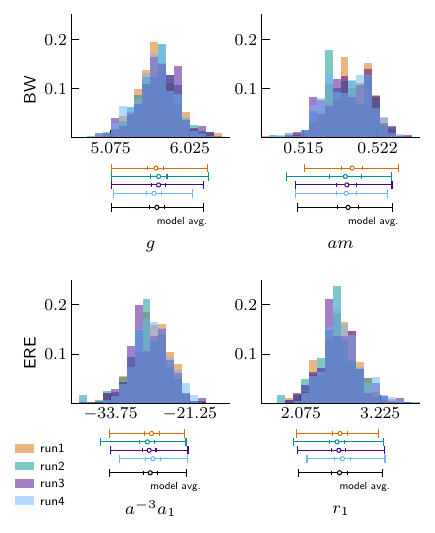}
    \caption{Weighted histograms displaying the distribution of the $K^*$ phase-shift fit parameters in lattice units are presented. The top row shows the Breit-Wigner parameters $g$ and $m$, while the bottom row illustrates the effective range expansion parameters $a_1$ and $r_1$. The four \textit{runs} represent variations in the fit ranges used for the underlying correlator fits, which are combined to obtain the final result. Further details can be found in~\citep{Boyle:2024grr}.\newline}
    \label{fig:histo-ps}
    \end{subfigure}
    \hspace{0.1cm}
    \begin{subfigure}{0.38\textwidth}
        \includegraphics[width=\textwidth]{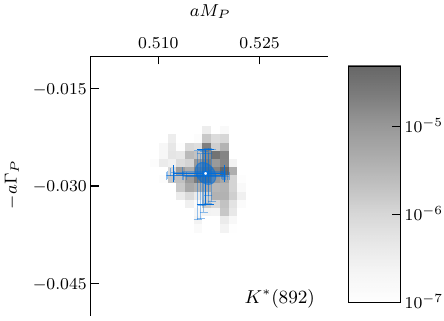} \\
            \includegraphics[width=\textwidth]{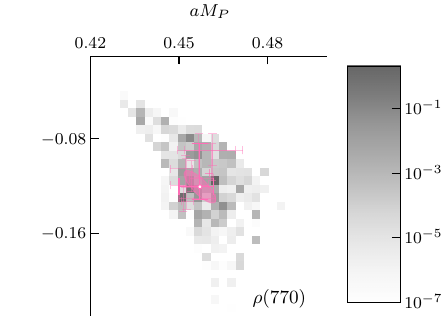}\\
            \vspace{2cm}
    \caption{2D histogram of the resonance pole positions in the complex plane, presented in lattice units, for both the $K^*$ and $\rho$ resonances. These positions are determined by solving Eq.~\eqref{eq:pole-continuation} using a phase shift parametrization derived from the parameters shown in Fig.~\ref{fig:histo-ps}. The grayscale shading represents the weighted frequency of each bin.}
    \label{fig:histo-poles}
    \end{subfigure}
    \caption{Histograms of the final results of the $K^*$ and $\rho$ scattering study. Figures taken from~\citep{Boyle:2024grr,Boyle:2024hvv}}
\end{figure}

Our solution involves importance sampling. For each energy level, we randomly draw a representative fit, with the likelihood of selection determined by $w_{\sf corr}$. A full set of these fits constitutes a \textit{collection sample} $s$. For each sample, we perform a phase-shift fit and compute an associated AIC score, $\textsf{AIC}_{\textsf{PS}}$. Repeating this process across multiple samples allows us to explore the fit space systematically. A histogram of phase shift parameters, such as the mass $m$ and coupling $g$ in a Breit-Wigner fit, is then produced. Since the underlying data was already weighted by $w_{\sf corr}$, applying an additional weight $w_{\sf PS}$ derived from $\textsf{AIC}_{\textsf{PS}}$ results in a total weight $\textsf{AIC}_{\textsf{tot}}$. In the infinite sampling limit, this approach converges to the ideal AIC score $\textsf{AIC}_{\textsf{ideal}}$. The histograms obtained for $K\pi$ scattering phase-shift parameters are shown in Fig.~\ref{fig:histo-ps}.
Resonance pole positions can then be determined by solving Eq.~\eqref{eq:pole-continuation} for the parameterized phase shift $\delta(\sqrt{s})$, as shown in Fig.~\ref{fig:histo-poles}. Since our study is limited to a single lattice spacing, the pole positions exhibit discretization effects. Ideally, these effects would be quantified by repeating the analysis on multiple lattice spacings. In the absence of such data, we estimate discretization effects conservatively using a power-counting argument. At approximately the 5\% level, these effects dominate the uncertainty in our final results. Our final results for the $K^*$ and $\rho$ pole positions show good agreement with experimental values~\cite{PDG2024}.

\section{Electroweak matrix elements}
\label{sec:eweak}

Lattice scattering computations, such as those described above, are valuable both for producing results testable against experiments and for predicting new QCD states. An especially intriguing application involves studying processes that involve an electroweak matrix element with multi-hadron final or intermediate states. The foundation for relating finite-volume and infinite-volume transition matrix elements originates from the work of Lellouch and Lüscher~\citep{Lellouch:2000pv}, who examined the hadronic $K\to\pi\pi$ decay. This formalism was later expanded to handle states with non-zero total momentum in finite volume, particles with intrinsic spin, and coupled two-particle channels involving non-identical and non-degenerate particles~\citep{Kim:2005gf,Christ:2005gi,Meyer:2011um,Hansen:2012tf,Briceno:2012yi}.

A further significant extension is the formalism for $1+\mathcal{J} \to 2$ transitions~\citep{Agadjanov:2014kha,Briceno:2014uqa,Briceno:2015csa}, describing processes where an external current is involved, enabling the extraction of resonance form factors. Frameworks for $2+\mathcal{J} \to 2$ transitions have been developed~\citep{Briceno:2015tza,Baroni:2018iau,Briceno:2020xxs}, although no numerical results have yet been published.

In the following, I will review the current progress in computations using these formalisms and discuss new studies presented at the conference.

\subsection{Hadronic decays}

One of the most detailed lattice QCD studies of multi-hadron weak decays is the $K\to\pi\pi$ calculation by the RBC/UKQCD collaboration~\citep{RBC:2020kdj}. This channel is particularly significant because it enables the determination of the direct kaon CP-violation parameter $\epsilon'$. In their most recent published result, they resolved a long-standing puzzle by including a lattice calculation directly at the physical pion mass. Experimentally, the ratio of the real parts of the $K\to\pi\pi$ amplitudes $A_I$ for the isospin channels $I=0,2$ is measured as $\mathrm{Re}(A_0)/\mathrm{Re}(A_2) = 22.45(6)$, with $A_0$ being significantly enhanced compared to $A_2$. This phenomenon is referred to as the $\Delta I=1/2$ rule and has long posed a challenge for perturbative approaches, which underestimate the ratio by approximately a factor of 10. The lattice computation revealed that $\mathrm{Re}(A_2)$ is highly sensitive to the light-quark mass due to significant cancellations of leading contributions at physical kinematics. Only by directly simulating at the physical pion mass, $M_\pi^\mathrm{phys}$, were they able to reproduce the experimental ratio, obtaining $\mathrm{Re}(A_0)/\mathrm{Re}(A_2) = 19.9(2.3)(4.4)$, thus providing an explanation for the discrepancy.

At this year's conference, two independent updates on the calculation of $\epsilon'$ were presented, both by the RBC/UKQCD collaboration. The first approach~\citep{Lattice24:Kelly} builds on the existing calculation by introducing a second, finer lattice spacing. It uses G-parity boundary conditions to avoid the need for explicitly removing certain multi-hadron states that could introduce growing exponential contributions. Ensemble generation for this study has progressed sufficiently for measurements to be underway, targeting a precision of 10\%.

The second approach~\citep{Lattice24:Tomii} uses periodic boundary conditions and employs the GEVP method, to determine the excited-state spectrum and account for them explicitly in the finite-volume formalism. Preliminary results were presented using two lattice spacings, with ongoing efforts to include a third, finer one.

\subsection{$1+\mathcal{J} \to 2$ transitions}

Finite-volume matrix elements describing a $1+\mathcal{J} \to 2$ transition can be expressed in terms of a "finite-volume form factor"~\citep{Briceno:2021xlc}
\begin{align}
  \langle n, {\bf P}_f | J^\mu | {\bf P}_i \rangle_L = \frac{1}{L^3} \frac{1}{\sqrt{2E_i}} \frac{1}{\sqrt{2E_n}} \mathcal{K}^\mu \mathcal{F}_L \, ,
\end{align}
where ${\bf P}_i$ and ${\bf P}_f$ are the initial and final momenta, and $\langle n, {\bf P}_f |$ represents an optimized resonance state obtained from a GEVP scattering analysis. $\mathcal{K}^\mu$ is a kinematic factor, and the finite-volume form factor $\mathcal{F}_L$ is related to its infinite-volume counterpart $\mathcal{F}$ through the Lellouch-L\"uscher factor.·

Initial applications of this formalism explored the $\pi\gamma \to \pi\pi$ transition~\citep{Briceno:2016kkp,Alexandrou:2018jbt}. More recently, a study of the $K\gamma \to K\pi$ transition appeared as well~\citep{Radhakrishnan:2022ubg}. At this year's lattice conference, progress on a study of nucleon transition matrix elements $\langle N\pi | J | N \rangle$ at a single lattice spacing with $M_\pi = 420$ MeV was presented~\citep{Barca:2024sub}. They presented an impressive improvement in matrix element extraction by using the GEVP method. This calculation is highly relevant for comparisons with experimental results from DUNE, Hyper-Kamiokande, and other experiments. Another independent effort~\citep{Lattice24:Gao} reported progress on the $N\gamma^* \to N\pi$ transition using two ensembles at the physical pion mass. Preliminary results were compared with ANL-Osaka experimental data.

\begin{figure}
    \centering
    \includegraphics[width=5cm]{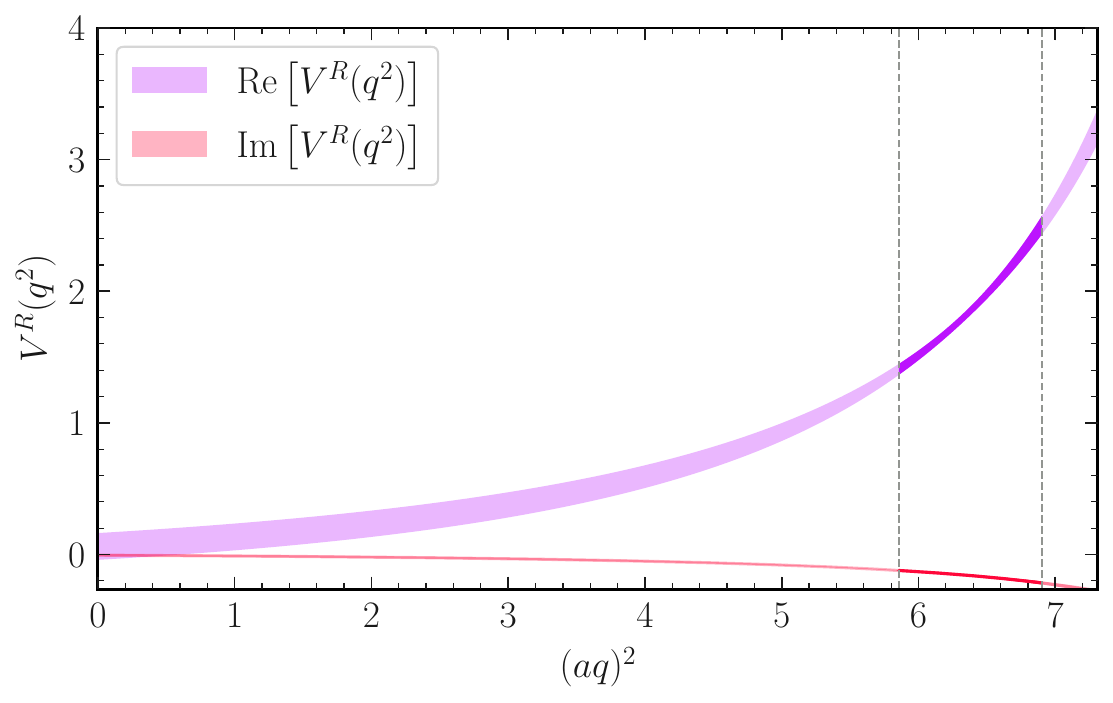}
    \includegraphics[width=5cm]{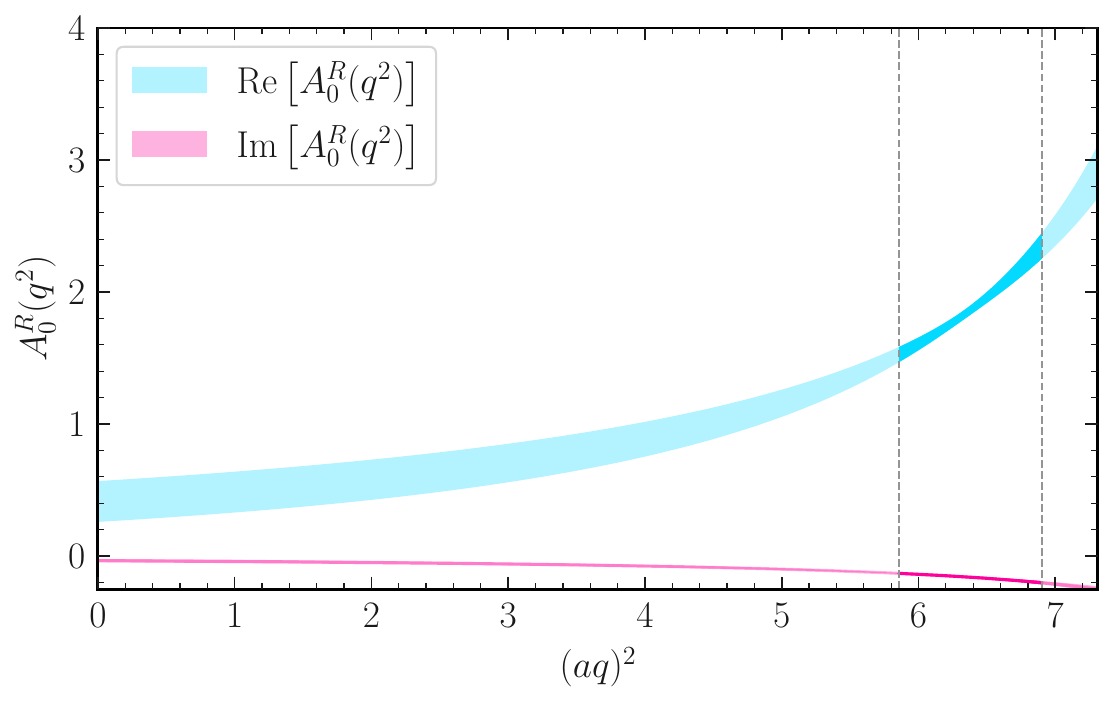}   
    \includegraphics[width=5cm]{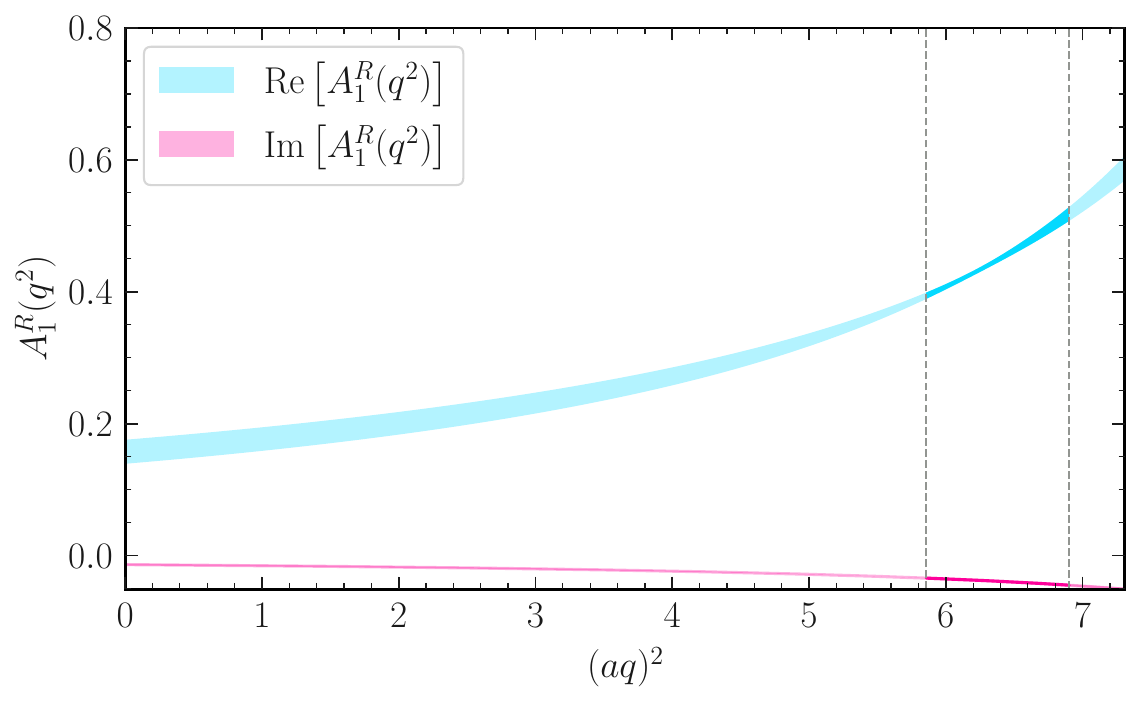}
    \includegraphics[width=5cm]{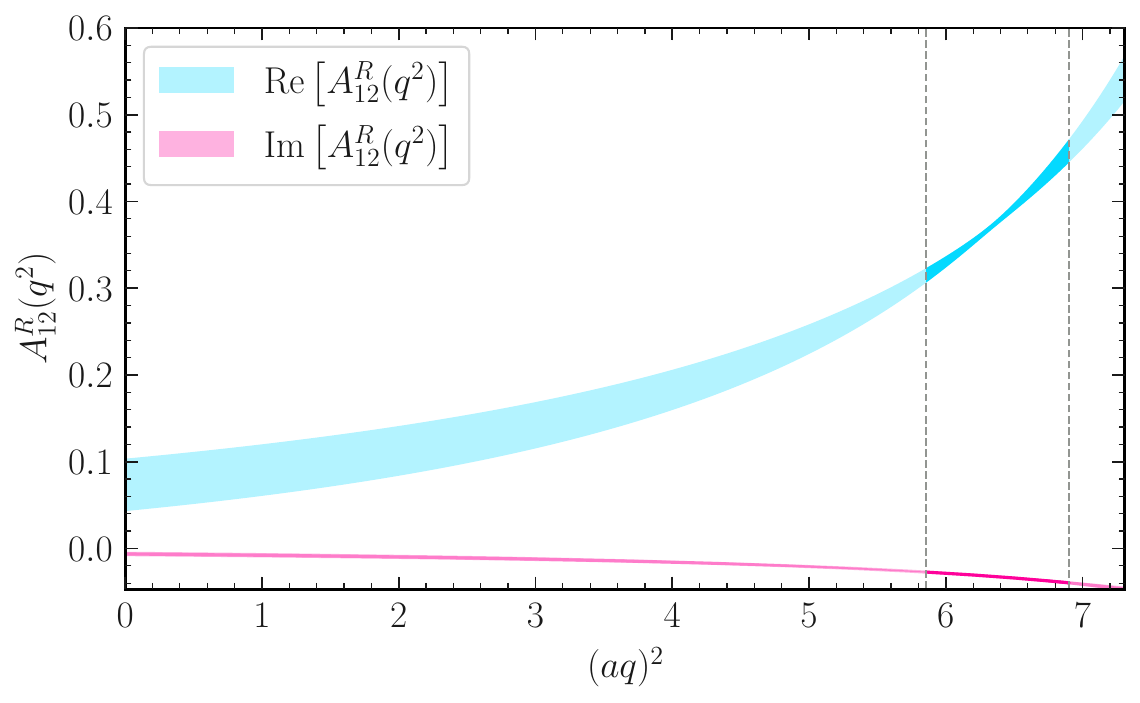}
    \caption{The four form factors describing the $B\to\rho\ell\nu$ transtion as shown in~\citep{Leskovec:2025gsw}. Lattice data is available at the high-$q^2$ region, between the black vertical lines. The lighter shaded region outside these lines is an extrapolation from the lattice data on the first-order $z$-expansion.}
    \label{fig:Brho}
\end{figure}

Progress has also been made on the $B\to\rho\ell\nu$ decay, with the calculation recently completed~\citep{Leskovec:2025gsw}. This channel requires extracting four form factors from vector and axial-vector currents:
\begin{align}
\langle n;\mathbf{p}_{\pi \pi} | J_V | B,\mathbf{p}_B \rangle_L &= \mathcal{C}_V^\mu  F_V(L) \, , \\
\langle n; \mathbf{p}_{\pi \pi} | J_A | B,\mathbf{p}_B \rangle_L &= \sum_{i=0}^2\mathcal{C}_{A_i}^\mu F_{A_i}(L) \, .
\end{align}
A significant challenge in this calculation arises from the lattice discretization of the heavy $b$-quark. This is addressed using an anisotropic Clover action which requires careful parameter tuning to minimize heavy-quark discretization errors. Further details on heavy-quark discretizations can be found in the quark flavour physics proceedings by J.T. Tsang~\citep{Lattice24:tsang}. This $B\to\rho\ell\nu$ study was performed at $M_\pi = 317$ MeV, a pion mass heavier than physical, but light enough for the $\rho$ to remain unstable and decay into two pions. All four form factors were extracted in the high-$q^2$ region where lattice data is available and are shown in Fig.~\ref{fig:Brho}.

This work opens the path towards computing semileptonic $B\to K^*\ell^+\ell^-$ decays, which require seven form factors to be described. These decays are crucial in the context of the so-called $B$ anomalies. Experimental results from the LHCb collaboration for $b\to s\ell^+\ell^-$ rare decays, such as $B\to K$~\citep{LHCb:2014cxe}, $B_s \to \phi$~\citep{LHCb:2021zwz}, and $B\to K^*$~\citep{LHCb:2016ykl}, consistently lie below theoretical predictions, especially in the low $q^2$ region which is less accessible for lattice computations and where predictions often rely on light-cone sum rules. A recent HPQCD result on $B\to K$ rare decay form factors~\citep{Parrott:2022zte}, extending into the low $q^2$ region, has reinforced the tensions with experimental data. Applying the techniques employed in~\citep{Leskovec:2025gsw} to the $B\to K^*$ decay channel could play a decisive role in investigating these anomalies and searching for potential signs of new physics.

\subsection{Long-range electroweak matrix elements}
Another important application is the extraction of long-range electroweak matrix elements, which requires a detailed understanding of strongly-coupled intermediate multi-hadron states. Extensions of the L\"uscher formalism were initially developed for rare kaon decays~\citep{Christ:2015aha} and kaon mixing~\citep{Christ:2015pwa}, later generalized for broader applications~\citep{Briceno:2019opb}, but also for explicit decays like the rare hyperon decay $\Sigma^+ \to p \ell^+ \ell^-$~\citep{Erben:2022tdu}. In this decay the lattice computation aims to extract the decay amplitude
\begin{align}
\mathcal{A}^{rs}_\mu = \int d^4 x \, \langle p,r | T [\mathcal{H}_W(x) J_\mu(0) ]| \Sigma^+ ,s\rangle \, ,
\label{eq:hyperon-amplitude}
\end{align}
involving a time-ordered product of the weak Hamiltonian $\mathcal{H}_W$ and the electromagnetic current $J_\mu$, with $r,s$ denoting the spinor indices of the baryon states.

The finite-volume estimator of this amplitude
\begin{align}
F_\mu(\mathbf{k},\mathbf{p})_L = \sum_n \, \frac{ C_{n, \mu} }{2 E_{n}(L) \big ( E_{\Sigma} - E_{n}(L) \big ) } + \dots
\end{align}
contains poles when a finite-volume energy state $E_n(L')$ equals $E_\Sigma$ for some $L'$. The full amplitude in Eq.~\eqref{eq:hyperon-amplitude} can be recovered from the finite-volume estimator by including an additive correction term
\begin{align}
\tilde{\mathcal{A}}_\mu(k,p)&=F_\mu(\mathbf{k},\mathbf{p})_L + \Delta F_\mu(\mathbf{k},\mathbf{p})_L \, , \\
\Delta F_\mu(\mathbf{k},\mathbf{p})_L &= i \mathcal{A}_{J_\mu}(E_\Sigma,\mathbf{k},\mathbf{p}) \mathcal{F}(E_\Sigma,\mathbf{k},L) \mathcal{A}_{H_W}(E_\Sigma,\mathbf{k}) \, ,
\end{align}
where $\Delta F_\mu$ depends on the $\Sigma \to N\pi$ amplitude $\mathcal{A}_{H_W}$, the $N\pi \to N\pi$ scattering amplitude, to which $\mathcal{F}$ is directly related, and the $p \to N\pi$ amplitude $\mathcal{A}_{J_\mu}$.

These methods have been successfully applied to rare kaon decays, first at $M_\pi \sim 430$ MeV~\citep{Christ:2016mmq} and later at the physical pion mass~\citep{RBC:2022ddw}. However, the latter calculation faced large errors due to cancellations in the GIM subtraction of up and charm quarks. Recent progress includes the application of the split-even approach~\citep{Giusti:2019kff,Harris:2023zsl} to rare kaon decays~\citep{Lattice24:hodgson}, showing error reductions of a factor of $4$ to $10$.

A recent calculation explored the long-distance contribution to the kaon-mixing CP-violation parameter $\epsilon_K$~\citep{Bai:2023lkr}. While experimentally measured with high precision, the theoretical determination is complicated by long-distance effects. This exploratory study, performed at unphysical light and charm quark masses, achieved a $40\%$ accuracy for $\epsilon_K$. However, the authors outline a path for future improvements, aiming to reduce the uncertainty to $10\%$ or lower.

Another investigation focused on the processes $B \to \mu^+\mu^-\gamma$, currently being searched for at LHCb~\citep{LHCb:2024uff}, and $B_s \to \phi\gamma$~\citep{Frezzotti:2024kqk}. The authors presented the vector, axial, and tensor form factors governing the transitions in the electroquenched approximation, while also suggesting methods to extend beyond this approximation. The phenomenologically significant charming penguin contributions were modeled as part of the study.

Furthermore, significant progress was reported on the first application of a novel formalism for computing the two-photon exchange contribution to $K_L \to \mu^+\mu^-$~\citep{Chao:2024vvl}. Initial results for the amplitude were obtained on an ensemble with near-physical pion mass and a coarse lattice spacing, with plans to extend the calculation using finer lattice spacings in the future.

\subsection{Long-distance contribution to the muon $g-2$}

\begin{figure}
     \begin{subfigure}{0.56\textwidth}
    \centering
    \includegraphics[width=\textwidth]{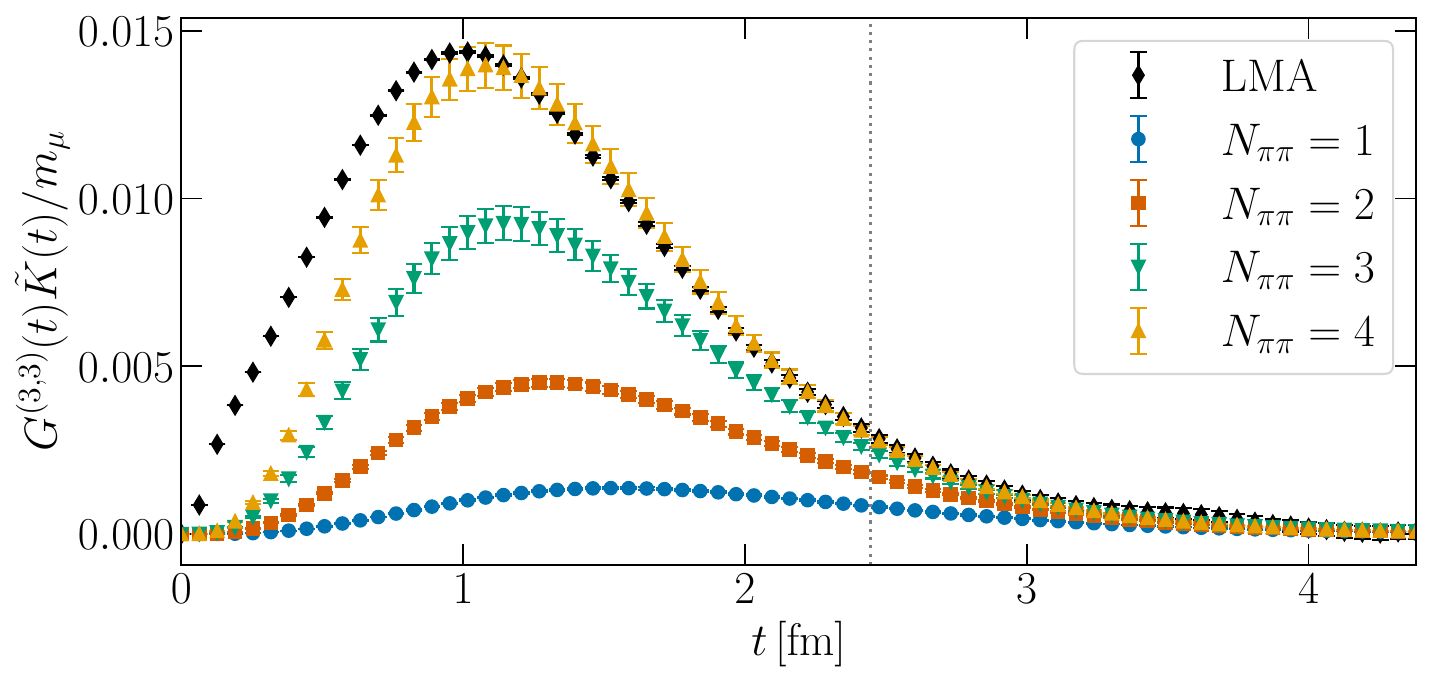}
    \caption{ }
    \label{fig:mainz-g2}
    \end{subfigure}
    \hspace{0.1cm}
    \begin{subfigure}{0.4\textwidth}
        \includegraphics[width=\textwidth]{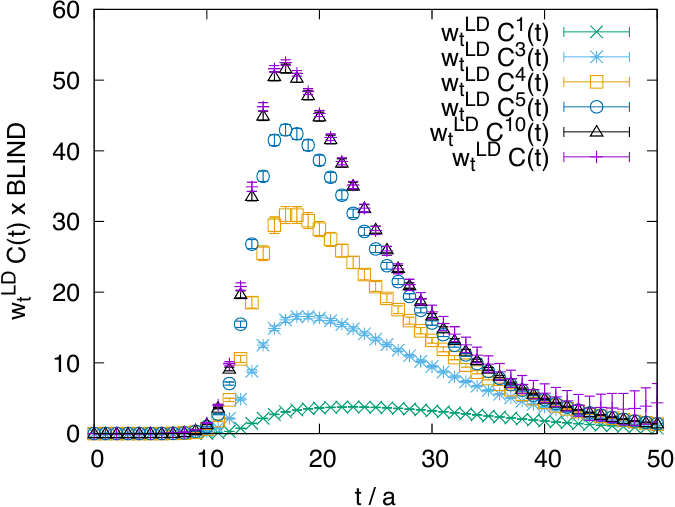} 
    \caption{ }
    \label{fig:RBC-g2}
    \end{subfigure}
    \caption{Plots displaying the reconstructed $g-2$ integrand from Mainz~\citep{Djukanovic:2024cmq} and RBC/UKQCD~\citep{RBC:2024fic}, both computed on ensembles with physical pion masses. In both cases, the reconstructed integrand closely matches the direct lattice calculation of the vector-vector correlator, shown as black (Mainz) and purple (RBC/UKQCD) data points. At large time separations, the reconstructed approach avoids the exponential signal-to-noise problem, yielding a highly precise estimator for the integrand.}
\end{figure}

Another important area where electroweak transition matrix elements involving scattering states have played a significant role is the muon $g-2$, discussed in detail in the plenary proceedings by C. Davies~\citep{Lattice24:davies}. This quantity is derived from an integral over the vector-vector correlator. Achieving the unprecedented precision required for $g-2$ calculations in lattice QCD involves different challenges across the short-distance, intermediate-distance, and long-distance components of the vector-vector correlator. Specifically, the long-distance contribution faces an exponential signal-to-noise problem, but it can be reconstructed from $\pi\pi$ scattering states~\citep{DellaMorte:2017khn,Bruno:2019nzm}
\begin{align}
G(t) = \sum_{n=0}^{N} \langle E_n, L | J(t) |0 \rangle e^{-E_n t} \, ,
\end{align}
where the finite-volume energy states $\langle E_n, L |$ are obtained from a GEVP analysis of $\rho \to \pi\pi$ scattering. This precise estimation method has been employed in the most recent long-distance $g-2$ calculations by the Mainz group~\citep{Djukanovic:2024cmq} (shown in Fig.~\ref{fig:mainz-g2}) and RBC/UKQCD~\citep{RBC:2024fic} (shown in Fig.~\ref{fig:RBC-g2}). Additionally, Mainz presented related progress on the extraction of the timelike pion form factor~\citep{Lattice24:Miller} at the physical pion mass.

\section{Outlook}
\label{sec:outlook}

A deep understanding of multi-hadron scattering is essential for lattice QCD computations aiming to provide rigorous tests of the Standard Model through comparison with experimental measurements. Two landmark examples of such calculations are hadronic $K \to \pi\pi$ decays and the muon $g-2$, both of which incorporate hadron scattering as a key component while also controlling the chiral-continuum extrapolation and presenting a complete error budget - a necessity for comparison with experimental results.

It is encouraging to witness both a surge in new scattering calculations and the continuous development and application of finite-volume formalisms, allowing a broader range of processes, including those involving three hadrons, to be treated rigorously. Additionally, an increasing number of computations are being performed directly at physical quark masses, a critical factor for drawing reliable conclusions from comparisons with experimental data.

As scattering computations reach this level of maturity, it will become increasingly important to establish quality criteria, similar to those employed by FLAG~\citep{FlavourLatticeAveragingGroupFLAG:2024oxs}, to ensure control over the underlying scattering states. With non-perturbative QCD results playing a growing role in precision physics and the lattice QCD community expanding the scope and applicability of scattering calculations, I am optimistic about the coming years, where lattice QCD will be a decisive tool in the search for physics beyond the Standard Model.

\acknowledgments

I extend my gratitude to the organizers of Lattice 2024 for the opportunity to present a plenary talk on scattering and its influence on flavour physics. I am deeply thankful to all my collaborators, many of whom have become close friends through years of shared work. Ambitious lattice QCD calculations are inherently collaborative endeavors, made possible only through collective effort and teamwork.

I especially thank Matteo di Carlo and Max Hansen for their critical review of an earlier version of this manuscript.

This work has received funding from the European Union’s Horizon Europe research and innovation programme under the Marie Sk\l{}odowska-Curie grant agreement No. 101106913.

\newpage

\bibliographystyle{JHEP}
\bibliography{proceedings}

\end{document}